# A Crossbar Network for Silicon Quantum Dot Qubits

R. Li[1,2], L. Petit[1,2], D.P. Franke[1,2], J.P. Dehollain[1,2], J. Helsen[1], M. Steudtner[3,1], N.K. Thomas[4], Z.R. Yoscovits[4], K.J. Singh[4], S. Wehner[1], L.M.K. Vandersypen[1,2,4], J.S. Clarke[4], and M. Veldhorst[1,2]*

[1]*QuTech, Delft University of Technology, P.O. Box 5046, 2600 GA Delft, The Netherlands.*
[2]*Kavli Institute of Nanoscience, Delft University of Technology, P.O. Box 5046, 2600 GA Delft, The Netherlands.*
[3]*Instituut-Lorentz, Universiteit Leiden, P.O. Box 9506, 2300 RA Leiden, The Netherlands.*
[4]*Components Research, Intel Corporation, 2501 NW 229th Ave, Hillsboro, OR 97124, USA.*
*email address: m.veldhorst@tudelft.nl

The spin states of single electrons in gate-defined quantum dots satisfy crucial requirements for a practical quantum computer. These include extremely long coherence times, high-fidelity quantum operation, and the ability to shuttle electrons as a mechanism for on-chip flying qubits. In order to increase the number of qubits to the thousands or millions of qubits needed for practical quantum information we present an architecture based on shared control and a scalable number of lines. Crucially, the control lines define the qubit grid, such that no local components are required. Our design enables qubit coupling beyond nearest neighbors, providing prospects for non-planar quantum error correction protocols. Fabrication is based on a three-layer design to define qubit and tunnel barrier gates. We show that a double stripline on top of the structure can drive high-fidelity single-qubit rotations. Qubit addressability and readout are enabled by self-aligned inhomogeneous magnetic fields induced by direct currents through superconducting gates. Qubit coupling is based on the exchange interaction, and we show that parallel two-qubit gates can be performed at the detuning noise insensitive point. While the architecture requires a high level of uniformity in the materials and critical dimensions to enable shared control, it stands out for its simplicity and provides prospects for large-scale quantum computation in the near future.

## I. INTRODUCTION

The widespread interest in quantum computing has motivated the development of conceptual architectures across a range of disciplines [1–5]. Effort to demonstrate the physical operation has culminated in the realization of high-fidelity single-qubit rotations, two-qubit logic gates, small quantum algorithms and simple quantum error correction schemes [6–13]. These confirm the suitability of several of these quantum systems on a single or few qubit level. The central next challenge is the scaling of qubit numbers so that practical computations can be performed [5,14,15]. Remarkable differences between the various approaches become apparent when considering the physical size of the qubit. A recent proposal for a microwave-trapped ion quantum computer with two billion qubits puts the required area to an astonishing size of more than $100 \times 100$ m$^2$ [16]. The same number of superconducting qubits is estimated to require an area of $5 \times 5$ m$^2$ [5]. Qubits defined by the spin states of semiconductor quantum dots, on the other hand, could fit in an area less than $5 \times 5$ mm$^2$. Clearly, small components can provide essential benefits in terms of scalability. However, as current qubit technology requires control lines for every qubit, a key challenge in each case is to avoid an interconnect bottleneck for full control of the dense qubit grid [17].

Conventional processors can have more than two billion transistors on a $21.5 \times 32.5$ mm$^2$ die [18,19]. Such a high packaging density crucially relies on a limited number of input-output connections (IO's). Transistor-to-IO ratios can be as high as $10^6$ [17,19] due to integration of so-called crossbar technology. Combinations of row and column lines enable the identification of unique points on a grid structure, providing a mechanism for large-scale parallel and rapid read-/write-instructions. In decades of advancements in semiconductor technology, this concept has resulted in today's most powerful supercomputers. Recently, the idea to implement similar shared control schemes for quantum systems has been recognized and proposed for semiconductor spins [17,20,21]. In one scheme, donor-based qubits are controlled via quantum dots at the crossing points of a large grid [20]. In a second scheme, quantum dot qubits are controlled by floating gates addressed via transistor circuits connected to a crossbar array [21]. This stimulated early proof-of-principle operations, such as local transistor-controlled charge detection [22], but does require extensive developments in down-scaling and developing new devices such as vertical transistors. Thus, while both proposals offer the prospect of a significant reduction in the number of connections to external control logic, they also rely on feature sizes and integration schemes that are not compatible with today's industry standards and that are far beyond current experimental capabilities.

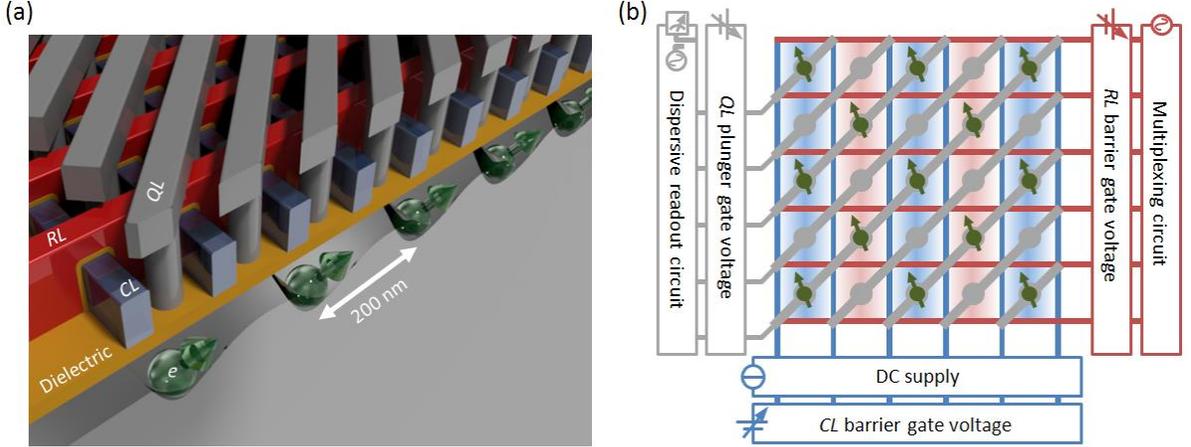

Fig. 1. Design of the quantum dot crossbar array. (a) 3D model of the array gate structure. The dielectrics in between the various gate layers are left out for clarity. (b) Schematic representation of the 2D quantum dot array. Column lines (*CL*, blue), row lines (*RL*, red), and qubit lines (*QL*, grey) connect the qubit grid to outside electronics for control and readout. A combination of these lines enables qubit selectivity. In the state shown here, half of the quantum dots are occupied with a single electron, where the electron spin encodes the qubit state. The electrons can be shuttled around via the gate voltages, providing a means to couple to nearest neighbors for two-qubit logic gates and readout, and to couple to remote qubits for long-range entanglement.

Here, we propose a crossbar scheme for a 2D array of quantum dots that can operate a large number of qubits with high fidelity. The structure is simple and elegant in design, is not requiring extremely small feature sizes, but instead relies on a high level of uniformity. Specifically, we require that a single voltage applied to a common gate can bring individual dots to the single-electron occupancy. In addition, depending on the operation mode, we require that the variation of tunnel coupling between quantum dots can be engineered to be within one order of magnitude. Continuous progress in fabrication have already led to individual double-dot systems with this level of charge uniformity [23–25]. We envisage that metrics such as variation in threshold voltage, charging energy, and tunnel coupling will need to improve by approximately an order of magnitude in order to use common gates in large quantum dot grids, and a promising platform to achieve this is advanced semiconductor manufacturing. Building upon such arrays we introduce a spin-qubit module that combines global charge control, local tunability and electron shuttling between dots with alternating local magnetic fields and global electron spin resonance (ESR) control. Truly large-scale quantum computing can be achieved by connecting multiple of these qubit modules. We will conclude by providing an overview of the challenges and opportunities for quantum algorithms and quantum error correction on the crossbar spin-qubit network.

## II. RESULTS

### A. Crossbar network layout

Figure 1 schematically shows the gate layout of the qubit module containing a two-dimensional (2D) quantum dot array. The qubits are based on the spin states of single electrons that are induced by electric gates in isotopically purified silicon ($^{28}$Si) quantum dots, reducing decoherence due to nuclear spin noise [26]. The architecture is agnostic to the integration scheme and the quantum dots can be located at a Si/SiO$_2$ interface [27], where the abrupt change in band structure can cause a large valley splitting energy leading to well-isolated qubit states [10,28]. Alternatively, the quantum dots can be formed in a Si/SiGe quantum well stack [29], where the epitaxial nature of the SiGe interface may be beneficial to meet the required uniformity for global operation as considered in the architecture here.

The architecture consists of a crossbar gate structure of three in-plane layers, see Fig. 1(a) and (b), and superconducting striplines on top. The striplines deliver global radio frequency (RF) pulses to manipulate the spin state, as will be discussed below. The first layer hosts the column lines (*CL*), which supply voltages to the horizontal barrier gates. The *CL*s also carry direct current (DC) for the generation of the magnetic field pattern [see also Fig. 2(c) and 2(d)]. These gates are deposited as the first layer to accommodate a well-defined cross-section and are made of superconducting material. The subsequent row lines (*RL*) are isolated from the first layer of gates and supply the voltages to the vertical barrier gates. The plunger gates are formed through vias that connect to the qubit lines (*QL*). Importantly, this gating scheme does not require smaller manufacturing elements than the quantum dots and the inter-dot tunnel barriers. Here, we consider barrier and plunger gate width of 30 nm and 40 nm, respectively, and quantum dot pitch spacing of 100 nm. These numbers enable more than a 1000 qubits to fit in an area smaller than $5 \times 5$ μm$^2$ (note that in our architecture half of the quantum dots host a qubit, increasing the area by a factor of two). Importantly, these dimensions are compatible with 3D XPoint technology and multiple patterning [30,31].

Fig. 1(b) shows a conceptual image of a qubit module. In the idle state, each qubit has four empty neighboring dots. This is achieved by setting the bias voltages applied to the diagonal qubit gates, alternating between accumulation and depletion mode. This sparse occupation has several advantages: it increases the number of control gates per qubit without changing the physical gating density, the sparsely spaced qubits reduce crosstalk, and the empty sites will enable the shuttling of qubits between different sites. The gate pattern allows for selective addressing of qubits with the combined operation of the different gate layers, as discussed below. For $N$ qubits occupying a square dot array, the combined control reduces the total number of gate lines to $N_{totalwires} \approx 4\sqrt{2N} + 1$. The analog control signals can be fed through the qubit network at the periphery and no additional control elements are needed within the grid. This allows for a dense packing of the quantum dots.

Since each gate is shared by a line of quantum dots, a high level of uniformity across the whole structure is required. These requirements can, however, be relaxed significantly when aiming for parallel qubit operation in a line-by-line manner. Here the long coherence times of silicon qubits become crucial [10]. We require that the tunnel coupling $t_0$ can be globally controlled to below 10 Hz in the off-state and in the range 10 - 100 GHz in the on-state, depending on the operation mode. The lower bound is set by the error threshold due to unwanted shuttling during a quantum algorithm. We note that while our architecture does not pose a theoretical upper bound to $t_0$, as arbitrarily large detuning $\varepsilon$ could be applied to the empty dots to suppress unwanted processes, very large $t_0$ will require impractically large voltages on the gates. Similarly, variations in the chemical potential energy $\Delta\mu$ could be overcome by applying an even larger detuning energy $\varepsilon$, together by exploiting the regime where the tunnel barriers can be pulsed on and off. However, we require $\Delta\mu < E_C$, where $E_C$ is the charging energy. This significantly reduces overhead in correcting pulses and pulsing amplitude, and increases operation speed (see the Supplementary Materials section 1 for details on uniformity and bounds).

Another challenge is to overcome cross talk, such that physical parameters as $\varepsilon$ and $t_0$ can be controlled individually [25]. Here, the highly repeatable nature and the presence of only straight lines in our architecture is strongly favorable. Compensating the crosstalk of an individual line by tuning the associated neighbor lines provides a highly symmetric approach. In the following discussion we assume the presence of such compensation, but refer to the main lines only.

### B. Magnetic field layout and ESR

Single qubit rotations are performed using global ESR striplines [see Fig. 2(a)] providing in-plane RF magnetic fields [10,32]. A modest external DC magnetic field is applied in the out-of-plane direction. Here, we consider an amplitude of ~ 3.6 mT, which corresponds to a resonance frequency $v_0$ ~ 100 MHz for the electron spin. This rather low magnetic field and resonance frequency eases the RF circuit design requirements. In addition, the qubit-to-qubit resonance frequency variation due to spin-orbit coupling [33–35] is strongly reduced in low magnetic fields and further minimized by applying the magnetic field perpendicular to the interface [36,37]. The ensemble ESR linewidth can then become narrow enough to achieve high-fidelity operation with a global ESR signal. Moreover, we expect improved qubit coherence due to a strongly reduced sensitivity to electrical noise in low fields, as coupling to charge noise via spin-orbit coupling is strongly reduced [36,37].

Local spin rotations could, in principle, also be implemented by integrating nanomagnets and operation based on electric dipole spin resonance (EDSR) [38]. To obtain Rabi frequencies $f_{Rabi}$ beyond 1 MHz, the required transverse field gradient is ~ 0.1-0.5 mT/nm for typical driving amplitudes and dot sizes [39,40]. However, while EDSR has proven powerful in single-qubit devices, the integration of nanomagnets in a dense 2D array is much more demanding. In particular, achieving the large required transverse field gradients will also lead to longitudinal field gradients. These will likely impact qubit coherence, shuttling and two-qubit logic gates. Furthermore, a large gradient appears incompatible with the low field operation proposed here. Therefore, qubit operation via ESR, requiring minimal field differences, is preferable for spin manipulation in this 2D array design.

To model the striplines and analyze the uniformity and amplitude of the RF fields generated by them, we use the Microwave Studio software package from Computer Simulation Technology (CST-MWS) [41]. This package has previously been used to accurately predict the RF magnetic field from transmission lines designed to drive single qubits [42]. To reach high uniformity across the 2D qubit array, we have designed a superconducting stripline pair. We use our CST-MWS model to optimize the relevant dimensions of the stripline design. Furthermore, to achieve homogenous fields the current distribution through the striplines has to be taken into account. For superconducting striplines, this is to a large extent determined by the superconducting penetration depth $\lambda$. In thin films with thickness $d$, the effective penetration depth is given by $\lambda_{eff} = \lambda_{bulk} \coth(d/\lambda_{bulk})$ [43]. As a result, $\lambda_{eff}$ in thin films can reach several micrometers, for example when using NbTiN with $\lambda_{bulk}$ close to 0.5 μm [44]. We find that already for $\lambda_{eff}$ > 0.5 μm, the corresponding RF field inhomogeneity across the 2D array can be less than $\delta v_1 = 2$ %, as shown in Fig. 2(b) (see the Supplementary Materials Section 3 for details). In addition, Rabi driving at 10 MHz requires 0.6 mA in each stripline and reasonable current densities $j_{Stripline} = 3 \times 10^9$ A/m$^2$ in the stripline pair for a thickness of 100 nm.

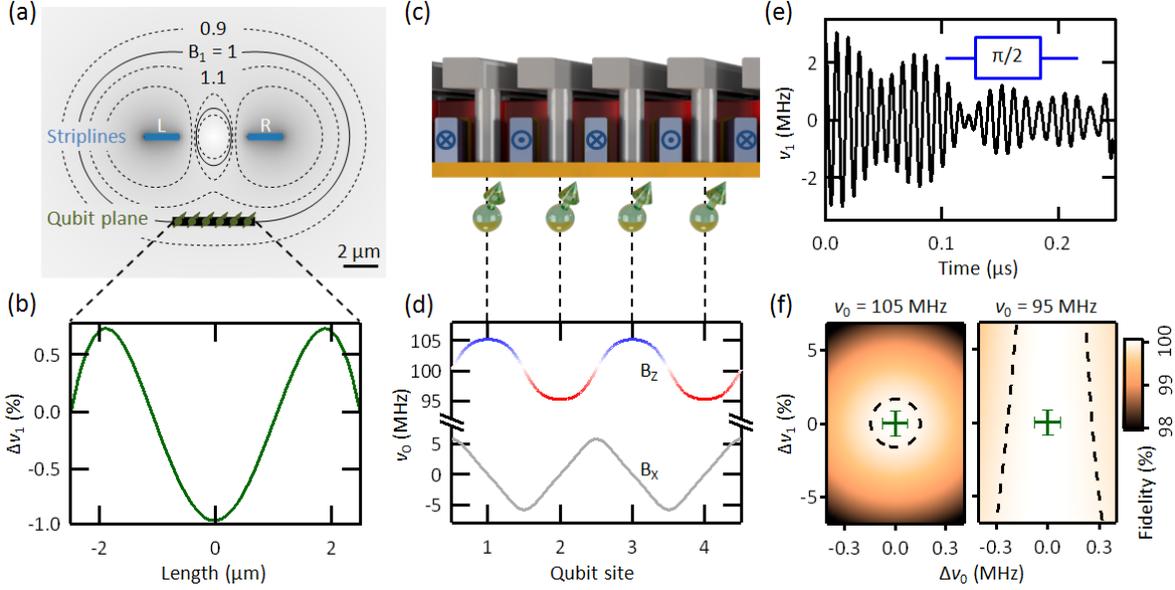

Fig. 2. ESR for single-qubit rotations and magnetic field profile of the crossbar structure. (a) Cross section of the stripline pair (2 μm wide and 6 μm pitch) positioned 4 μm above the qubit plane. The grey background with black contour lines visualizes the RF magnetic field generated by driving currents through the striplines. (b) The double stripline is optimized to minimize the variations in RF magnetic field at the qubit plane, and we find peak-to-peak values below 2%. (c) A DC alternating in direction between even and odd *CLs* together with an external out-of-plane field generates the static field profile shown in (d). The field component $B_Z$ has local maxima and minima at the qubit sites, where $B_x$ vanishes, providing qubit addressability and to first-order insensitivity to qubit placement. (e) Column-selective qubit pulses are engineered using GRAPE and here a selective π/2 pulse is shown, with fidelities shown in (f). The GRAPE pulse is designed to tolerate static variations in the $v_0$ and $v_1$. The green error bars denote the expected qubit-to-qubit variations taking into account the design considerations. We conclude that single-qubit rotations can be performed with fidelity higher than 99.9% in the 2D qubit array. Here, we use an electron *g*-factor of 2 and show the static field by the resonance frequency $v_0$ and the ESR field by the normal on resonance Rabi frequency $v_1$.

To achieve qubit addressability, a column-by-column alternating magnetic field is generated by passing direct currents with alternating directions through the *CL*, as shown in Fig. 2(c) and 2(e) (see also the Supplementary Materials Section 2). The targeted $\delta v_{CL} = 10$ MHz frequency difference between columns requires significant current densities $j_{CL} = 4 \times 10^{10}$ A/m$^2$ in the gate lines. Nonetheless, these current densities are below the superconducting critical current density of for example NbN [45]. The integration of superconducting lines suppresses heat dissipation. In addition, it minimizes potential differences along the lines. The expected field profile along a row of qubits is plotted in Fig. 2(d).

Spin-orbit coupling in silicon is strongly enhanced close to an interface and in the presence of large vertical electrical fields [46]. This can lead to significant qubit-to-qubit variations in resonance frequency [33–35]. These variations depend on the microscopic interface and even a single atomic step edge can have a strong impact; it will thus be a significant challenge to overcome these variations by fabrication methods only. In typical silicon metal–oxide–semiconductor (MOS) quantum dots the variations in the *g*-factor are up to $\Delta g/g = 1 \times 10^{-2}$ [33,34]. In SiGe devices, the variations are predicted to be an order of magnitude smaller, $\Delta g/g = 1 \times 10^{-3}$ [35]. Possible optimization strategies to reduce variations could focus on the perpendicular electric field, or on the materials stack. However, by operating in the low magnetic field regime and by applying the field perpendicular to the interface [36,37] as proposed here, the qubit-to-qubit variation is expected to vanish and we take a conservative estimate $\delta v_{SOC} = 50$ kHz.

Imperfect device fabrication can result in local variations of the magnetic field. This impact is minimized because the magnetic field is self-aligned with the quantum dot barriers defined by the *CLs*. Furthermore, the magnetic field pattern is designed to have local minima or maxima at the qubit positions, such that the qubit energy splittings are to first order insensitive to variations in location (see the Supplementary Materials Section 2 for more details). The dominant contributions to variations in $v_0$ will thus come from variations in the geometry of the gates. For a 1 nm root-mean-square (rms) variation in gate geometry we estimate the corresponding resonance frequency linewidth to be $\delta v_{fab}$ = 100 kHz. Based on these considerations we find a total variation $\delta v_0 = \delta v_{fab} + \delta v_{SOC}$ = 150 kHz.

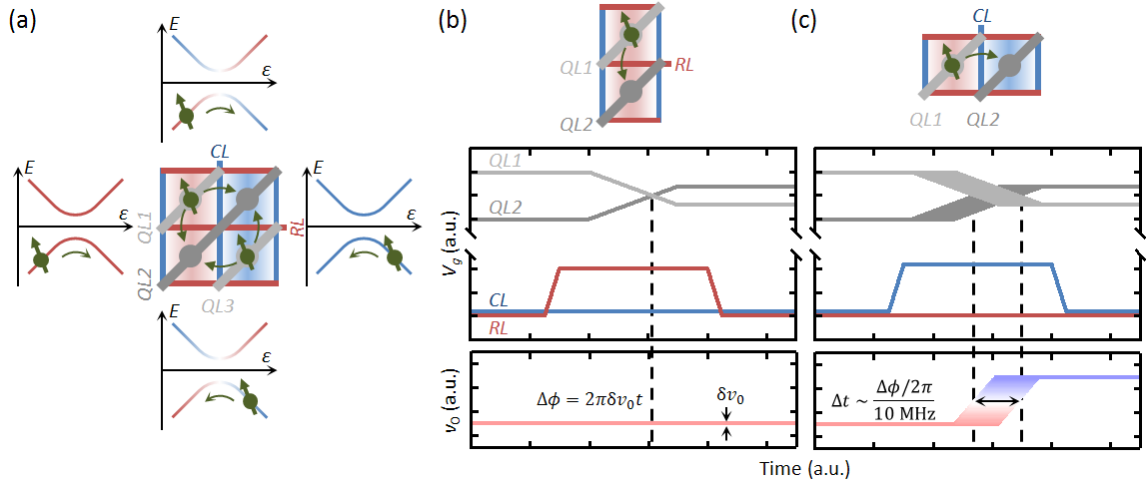

Fig. 3. Qubit shuttling in the crossbar array. (a) By controlling the tunnel coupling and potentials of the dots, qubits can be shuttled around. (b) Shuttling along a column. The sequence consists of setting the tunnel coupling by *RL* followed by pulsing the detuning energy. This process leaves the qubit resonance frequency unaffected except for unintended qubit-to-qubit variations. (c) Shuttling along a row. This process results in an additional 10 MHz shift, which is the basis of phase updates and *Z*-gates.

For the implementation of global high-fidelity single qubit operations it is central that the RF pulses are forgiving with respect to the inhomogeneity in field as discussed above. At the same time the pulses need to be highly frequency selective to ensure that no unintended qubit rotations or phase shifts are induced in the off-resonant columns. Considering these challenges, we applied Gradient Ascent Pulse Engineering (GRAPE) for ESR spin control [47], as shown Fig. 2(c). With this technique we can achieve single qubit fidelities above 99.9 % and crosstalk below 0.1 % and perform a $\pi/2$ rotation within 250 ns. The tolerance levels for this fidelity are up to 300 kHz in $v_0$ and over 3 % in $v_1$, indicated by the black dashed lines in Fig. 2(f). For comparison, we also include (green error bars) the expected qubit-to-qubit variation based on the discussion above, which falls well within the 99.9 % fidelity domain. We note that the error bars denote the peak-to-peak variations, such that many qubits will have significantly higher fidelity. This implies that further optimization could be done if a certain number of faulty qubits can be tolerated.

### C. Shuttling qubits for addressability and (long-range) entanglement

We now turn to the shuttling of electrons [48–51] as a means to create addressability for single and two-qubit logic gates, as well as an efficient method for (remote) qubit swap. The general principle behind the crossbar operation is the combined control of $\varepsilon$ and $t_0$. Since detuning and tunneling are controlled by different layers of gates, each qubit can be selectively addressed at the corresponding crossing point.

Figure 3 visualizes qubit shuttling along a row or column. Shuttling involves a change in the qubit resonance frequency. Therefore, the electron wave function has to be shifted diabatically with respect to the spin Hamiltonian, so that we can shuttle the qubit between different sites while preserving its spin state. By utilizing a non-linear pulsing scheme, we can operate the qubit shuttling up to at least 1 GHz with fidelity higher than 99.9% when accounting for small $t_0$ and large pulsing amplitude for uniformity requirements (see the Supplementary Materials Section 5).

The difference in Larmor frequency between adjacent columns can be exploited to construct fast *Z*-gates operating at 10 MHz, see Fig. 3(c). This can be utilized to correct phase errors or to implement a *Z*-gate in a quantum algorithm simply by temporarily moving a qubit to an adjacent column for a properly calibrated duration.

### D. Two qubit logic gates and Pauli spin blockade based readout

Two types of two-qubit gates can be implemented with quantum dots, namely the $\sqrt{\text{SWAP}}$ and the Controlled-Phase (CPhase) gate [11–13,52–56]. A direct implementation of the CPhase gate, however, requires the Zeeman energy difference to be much larger than the exchange coupling, $\delta E_Z \gg J$, in order to reach high-fidelity. The small field gradient $\delta E_Z = 10$ MHz considered here will not fully suppress SWAP-type rotations reducing the fidelity. A possible solution could be to engineer composite pulses, but here we focus on √SWAP as the central two-qubit gate, see Fig. 4(a). Together with single-qubit rotations this provides a universal quantum gate set. For example, a CNOT is obtained by interleaving a *Z*-gate in between two √SWAP operations, where the *Z*-gate can be conveniently realized by utilizing the shuttle scheme. In order to execute the √SWAP, we shuttle two qubits into the same column such that the *g*-factor difference is minimized, and we tune the qubit exchange by controlling the tunneling barrier gate while keeping the two qubits at the charge symmetry point with the qubit gates [55,56].

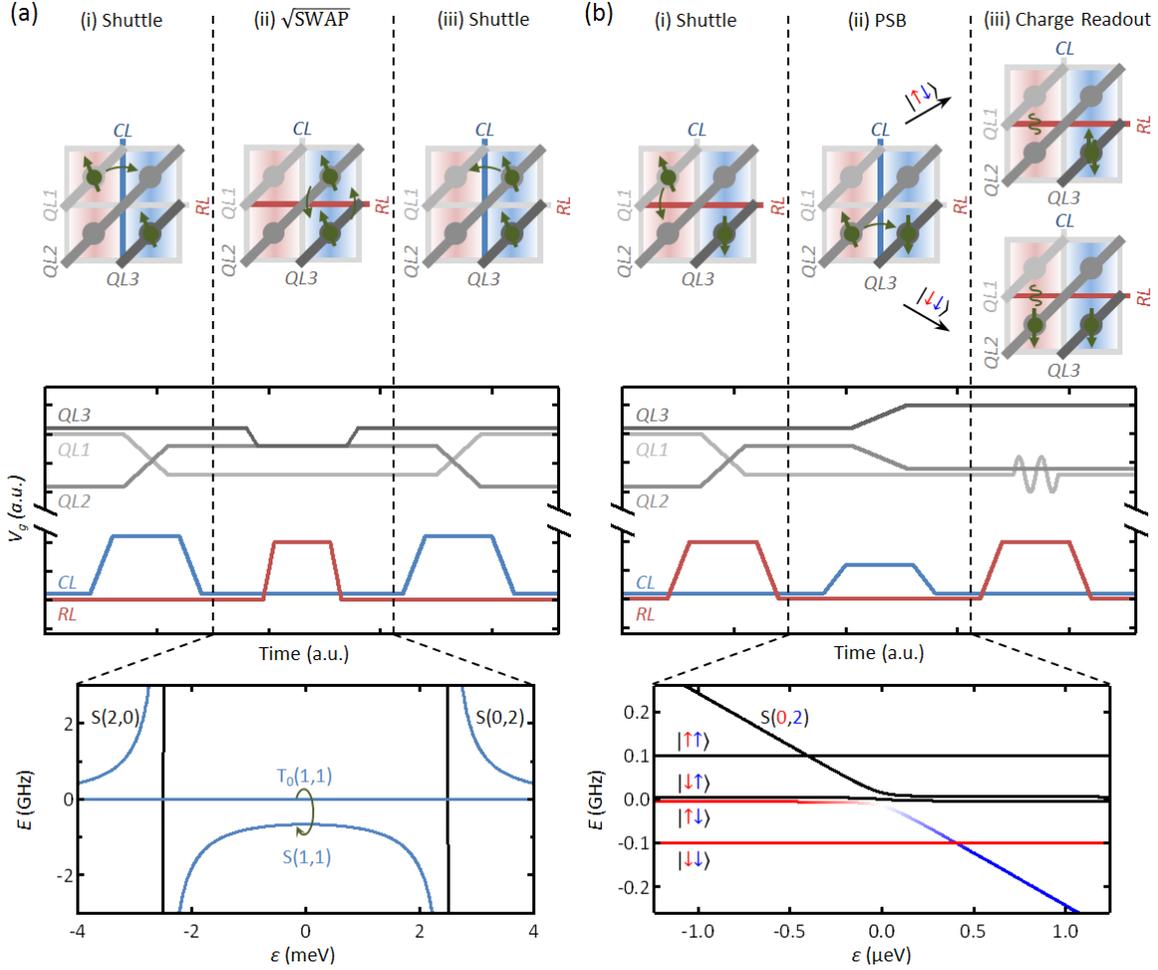

Fig. 4. Two-qubit logic gates and readout. (a) Sequence for √SWAP gates. By shuttling the respective qubits to the same column, the resonance frequency difference is minimized, enabling a high-fidelity √SWAP. The logic gate is performed at the symmetry point, making the qubits to first-order insensitive to detuning noise, and the interaction is controlled by the associated *RL*. (b) Spin qubit readout. Here, the respective qubits are shuttled to reside in the same row. The ancillary qubit, located at the blue column with the larger Zeeman energy, is manipulated to the spin down state. The measurement qubit is adiabatically pulsed. The qubit shuttles when the state is spin up and is blocked when the state is spin down, because of Pauli spin blockade. Subsequently, the tunnel coupling is turned off and the charge is locked. Dispersive charge state readout occurs by exploiting an empty neighbor dot.

In the low magnetic field regime discussed here, reservoir-based spin initialization and readout is not possible due to thermal broadening. Therefore, we utilize the Pauli spin blockade (PSB) between two electrons on neighboring sites for spin initialization and readout. This method has the additional advantage of not requiring a reservoir next to the qubit. The protocol relies on the difference in Zeeman energy between the two quantum dots to enable spin parity projection. This difference in energy is created by the same column-by-column alternating magnetic field used to create qubit addressability, and readout is performed between neighboring quantum dots in different columns.

The PSB spin-to-charge conversion scheme is plotted in Fig. 4(b). Instead of shuttling along a row, which brings two qubits to adjacent sites in the same column (same resonance frequency), the qubit is now moved along a column. This brings it next to a qubit in a different column, providing the difference in Zeeman energy that is necessary for readout. In the sequence shown here, the qubit with the smaller Zeeman energy (red background) will be read out. The qubit with the larger Zeeman energy (blue background) serves as an ancillary qubit and must be in the spin down state, which can be controlled via qubit pulses or leaving the qubit idle. Readout is achieved by pulsing towards a configuration where two electrons are favored on a single dot. Dependent on the spin state, PSB will or will not prevent one electron from moving over and joining the other. The above process completes the spin-to-charge conversion, and the spin state can be inferred from the charge occupation. A conversion fidelity higher than 99.9 % can be achieved with a 3 MHz gate pulsing speed [57] (see the Supplementary Materials Section 5). We note that in another protocol the ancillary qubit can be in the spin up state, provided it resides in the column with the smaller Zeeman energy (see the Supplementary Materials Section 6). This possibility could prove to be powerful in quantum error correction cycles, as it avoids the need to actively correct errors.

Directly after the PSB spin-to-charge conversion, we switch off the inter-dot tunnel coupling with *CL*, so that the charge state is disconnected from the spin configuration. In this mode the state is not sensitive to spin relaxation, thereby increasing the readout fidelity [57]. This can be exploited for delayed readout schemes, such as charge sensor based readout by shuttling to the periphery of the 2D array. However, here we consider gate-based dispersive readout [22,58,59] for an on-site readout of the charge state, as shown in Fig. 4(b). By applying an RF carrier signal to the qubit gates and coupling the dot to an adjacent empty dot, the charge state can be extracted from the dispersive signal. When there is charge occupation, the inter-dot oscillation driven by the RF carrier gives an additional quantum capacitance, leading to a different reflected signal compared to the state without charge occupation. By measuring the reflected signal we thus determine the qubit state.

### E. Parallel operation

For an efficient quantum computing scheme, simultaneous operation is essential. Here we discuss how the local operations introduced above can be advanced towards line-by-line or even near-global operation. Contrary to local operations, parallel operations result in active gates crossing at quantum dots that are not targeted, see Fig. 5. This may lead to undesired operations. However, these can be prevented by selectively occupying the quantum dots and specific control of $\varepsilon$ or $t_0$, such that away from the targeted locations signals are only applied to empty quantum dots or to quantum dots with empty neighbors.

Figure 5(a) shows an example of a line-by-line operation of controlled-phase shuttling. To properly control the timing it is crucial to individually pulse the *QL*. Still, parallel shuttling operations can be implemented along one column or row, enabled by lifting the barriers controlled by one *CL* or *RL*, respectively. These *CL*s can be time-controlled individually to correct the qubit-to-qubit variations, such that the shuttled qubits have the correct phase after the shuttling. The line-by-line shuttle can be performed within 1 ns with fidelity beyond 99.9% (see the Supplementary Materials Section 5).

An approach to performing simultaneous two-qubit logic operations on the qubit module could be to shuttle line-by-line all target qubits to the associated control qubits and then perform $\sqrt{\text{SWAP}}$ operations line-by-line. However, this will lead to qubit configurations where targeted qubits share gate lines disabling individual gate control, which is essential for high-fidelity operation. To overcome this, we propose sequences whereby a single column (or row) of qubits is shuttled first, followed by the desired operation and shuttle back, and then the sequence is continued by operating the next line of qubits of the module until all qubits are addressed. This protocol is demonstrated in Fig. 5(b), which shows the configuration after shuttling a single column of qubits. Now, targeted pairs of qubits can be tuned individually to their optimal configuration, by optimizing $\varepsilon$ and $t_0$. For example, operations can be performed at the detuning-noise insensitive charge symmetry point [55,56]. Consequently, the operation speed is not limited by the line-by-line control and we envision operation frequencies in the range 10 – 100 MHz for two-qubit logic gates.

Simultaneous readout consists spin-to-charge conversion step and charge readout step. First a row of qubits is shuttled, resulting in the configuration shown in Fig. 5(c). After that, the parameters $\varepsilon$ and $t_0$ can be individually controlled to convert spin-to-charge. In this specific sequence here, qubits are alternately shuttles up or down along the row, which leads to a configuration that is typically compatible with error correction sequences [5,60]. However, there may be instances where a different configuration is required, and this could reduce the spin-to-charge conversion to half the speed compared to line-by-line.

Near global operation is possible when phase control is not required. This may have multiple applications, for example in achieving long-range coupling. In such protocols, multiple shuttles can be performed with a single phase match at the start or at an arbitrary point. An example of global shuttling is shown in Fig. 6(a), where half of the qubits are simultaneously moved. Shuttling requires adiabatic movement with respect to the tunnel coupling and the demand is most stringent close to the anticrossing point. Due to qubit-to-qubit variations it may not be possible to go beyond a linear detuning pulse, as each pair can have the anticrossing at a different location. This consequently limits the shuttle speed. Nonetheless, for a $\Delta\mu = 2$ meV, shuttling can be at a 1 GHz rate when $t_0 > 25$ GHz (see the Supplementary Materials, section 5). This simultaneous shuttling can be highly important for advanced error correction codes that require long-distance coupling, such as the 3D gauge color code [61].

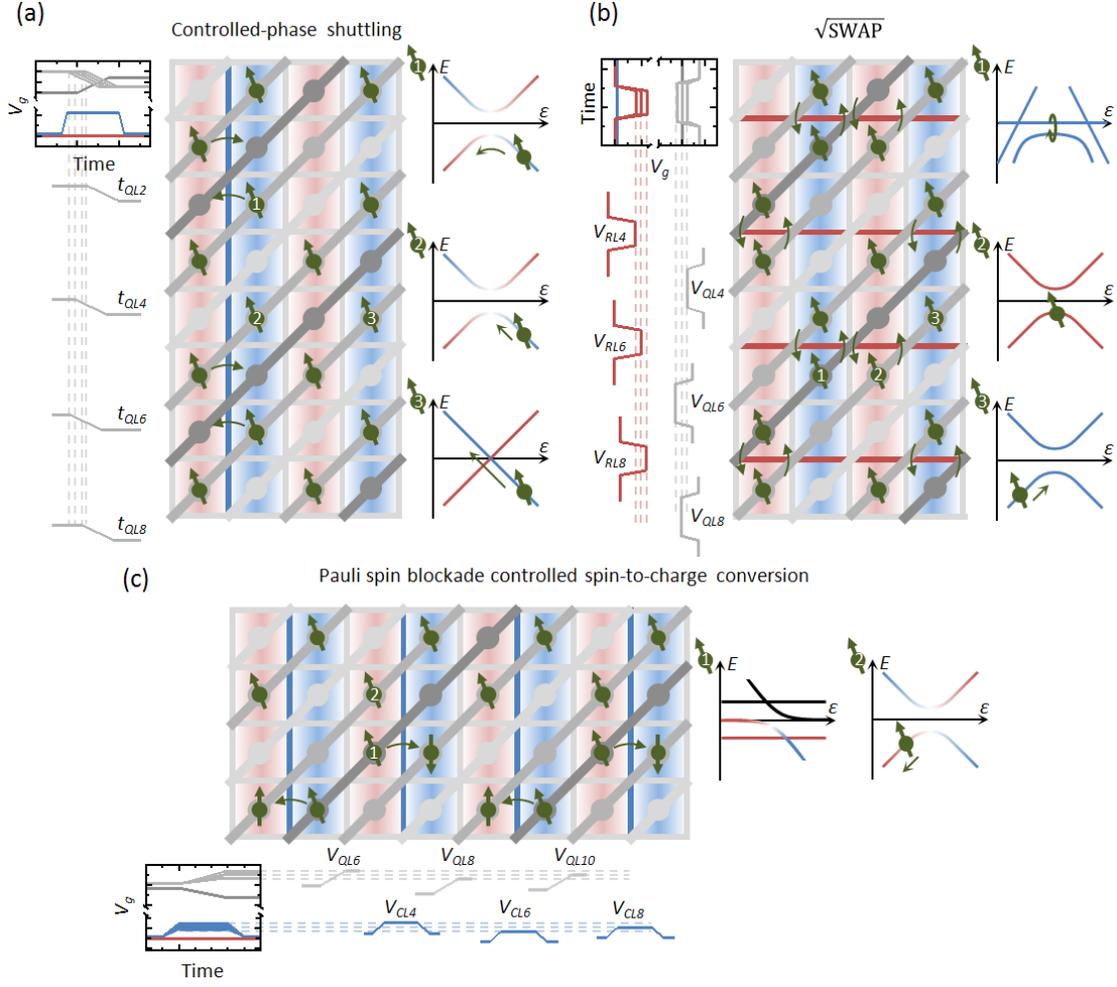

Fig. 5. Line-by-line operation. Simultaneous operation of controlled-phase shuttling (a), two-qubit √SWAP operations (b), and spin-to-charge conversion (c) can be achieved in a line-by-line manner. In each figure, inset (1) denotes the energy-detuning diagram of the targeted qubit(s). Inset (2) and (3) shows the consequence on the remaining qubits, where detuning, tunnel coupling, or the local magnetic field minimizes errors. (a) Shuttling of qubits. Parallelism is obtained along one direction and tunability along another direction, and the respective gates control the timing and detuning to overcome qubit-to-qubit variations. Here, the target qubits shuttle from column to column, whereas the other qubits are blocked by $\varepsilon$ or $t_0$. (b) Two-qubit logic gates. √SWAP operations only occur between tunnel-coupled neighboring qubits. The remaining qubits do not interact, but could shuttle in a column. The resulting (small) phase shift can be corrected by the consecutive shuttle event in the line-by-line operation. In (c) PSB spin-to-charge conversion occurs between tunnel-coupled qubits. Qubits coupled to an empty dot do not shuttle, prevented by the energy alignment, since we require $\Delta\mu < E_C$.

Global charge readout requires to distinguish between qubits connected to the same $QL$. This is achieved via frequency multiplexing. Here, an additional voltage modulation is applied to the $RL$, as shown in Fig. 6(b). The separation of spin-to-charge conversion and charge readout in different steps has a particular advantage. While the initial spin-to-charge conversion must be performed line-by-line it can be done relatively fast. The readout of charge is likely slower and to overcome the non-uniformity in $\Delta\mu$ a large detuning has to be applied. Instead of a single-step readout, we sequentially readout for different detuning and group the qubits according to their detuning (see the Supplementary Materials, section 5). This sequential readout as compared to the line-by-line approach has the advantage that is independent on the number of qubits and will be efficient for large qubit modules. The total readout time will strongly depend on the performance of dispersive readout at the single-qubit level, now under intensive research. However, the protocol here shows that the slowdown with increasing numbers of qubits can be controlled.

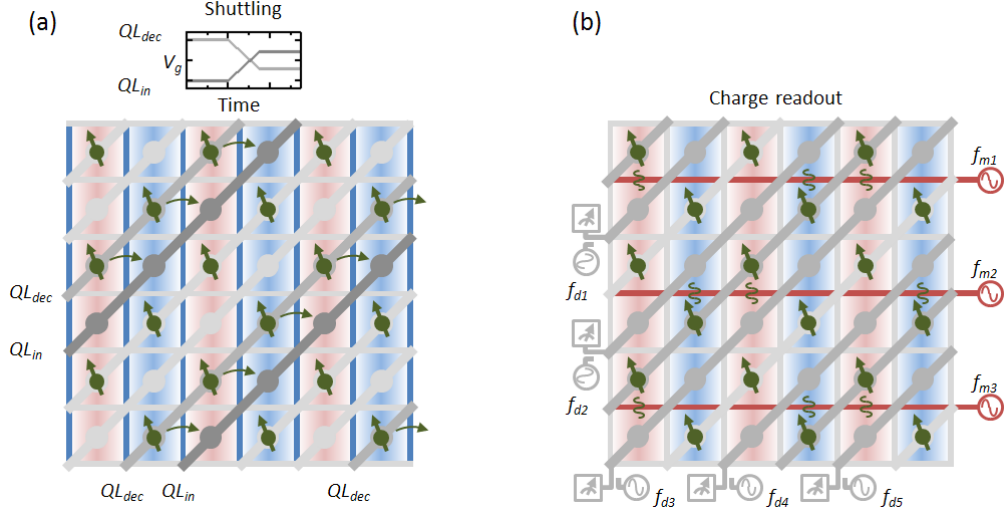

Fig. 6. Near global operation. Shuttling (a) and charge readout (b) can be performed in a near global manner. (a) Shuttling without phase control enables to construct a variety of shuttle patterns that can be operated almost globally; the schematic here shows the simultaneous shuttling of half of the qubits one site to the right. (b) The dispersive charge readout, performed after the spin-to-charge conversion shown in Fig. 5(c) can be performed simultaneously by including frequency multiplexing. The RF carrier on $QL$ ($f_d$) is then modulated by the application of additional multiplexing RF pulses ($f_m$) to $RL$.

### F. A network of qubit modules

For truly large-scale quantum computation, we envision a network consisting of a large number of interconnected qubit modules. While the layout of such an architecture crucially depends on the specific qubit module implementation and therefore goes beyond the current proposal, a possible repeatable tile is depicted in Fig. 7 (see also the Supplementary Materials, section 7). In addition to the central array hosting the qubit module, the quantum dot grid is extended in a simpler structure consisting of barrier gates only, thereby strongly reducing the number of required control lines. These shuttling dots cannot be fully controlled, but do allow for the transportation of qubits [48–50,62]. With this approach, qubit modules can then be connected together, where the available space can be used for local electronics [17,21,22] or wiring fan-out. Transportation of, e.g., a column of qubits from the edge of one module to another module would then provide a large range of possibilities for quantum algorithms, since it would create a large virtual array of coupled qubits with a certain degree of long-range coupling.

### III. DISCUSSION

One of the greatest challenges in the area of scalability is avoiding an interconnect bottleneck. Here we have proposed a scalable solution for spin qubits based on crossbar technology. While this technology limits control, we have developed general operation schemes sequences based on partial sequential control. The increased operation time due to sequential control is warranted by the very long coherence times of quantum dot spin qubits, with experimental demonstrations already up to 28 ms [10]. We have shown operation schemes for phase-controlled shuttling, two-qubit logic gates, and spin-to-charge conversion. These operations can have a targeted execution time well below 1 μs. The resulting loss of coherence due to the waiting time when operating in a line-by-line manner could be well below $10^{-3}$ in a 1000-qubit module using suitable echo sequences. The shuttling proposed here can be performed simultaneously within 1 ns, enabling even more than $10^7$ operations, and could provide an excellent method to create long-range entanglement or remote qubit SWAP. Readout could become fast by global operation, and measurement-free quantum error correction schemes could reduce the need for frequent readout [63,64]. If advances in qubit control continue to improve and lead to all fidelities greater than 99.9%, the architecture discussed here provides an excellent way forward to large-scale quantum computation.

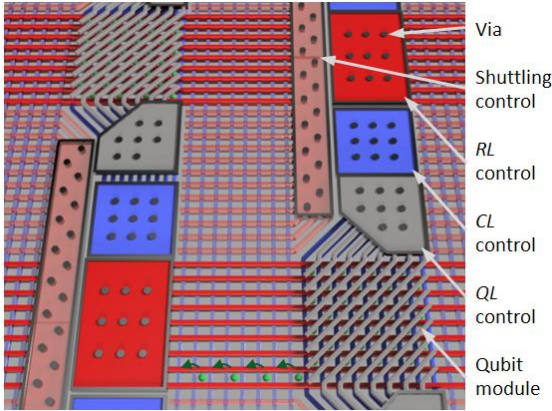

Fig. 7. Prospects for connecting qubit modules. Individual qubit modules (targeted to be of order 1000 qubits, for clarity reasons smaller modules are shown here) are connected together using long-range shuttle highways. The parallelism of the long-range shuttlers strongly reduces the number of control lines, providing space to integrate local electronics or vertical vias to interconnect the qubit array to outside electronics. Individual qubit modules could be operated using specific codes or be programmed to host, for example, a single logical qubit.

The proposed architecture supports universal quantum computation in a fault-tolerant manner [60], where the ability to shuttle qubits over large distances in principle provides means to realize quantum error correction schemes and quantum circuit implementations otherwise reserved for non-planar architectures. Within one qubit module, the highly flexible nature of the presented architecture makes it amenable to the use of a variety of topological error correction codes [60]. For planar codes, this includes the surface code [5], which has a fault tolerance threshold as high as 1% [65] and moreover can be implemented using entangling gates between qubits that are adjacent on a 2D surface. A distance three surface code would fit in a 7 × 7 quantum dot module and a successful implementation would present a milestone on the path towards fault tolerant quantum computation.

The proposed architecture is also amenable to other 2D local topological error codes such as the 2D color code [61], which has a lower threshold [66] but supports a more expansive set of logical operations. Finally we also envision that the use of qubit shuttling will enable the implementation of error correction schemes requiring long range entanglement such as the 3D gauge color code [61]. This approach has several highly desirable properties including low stabilizer generator weight, the possibility of a high error threshold [61] and the ability to perform (through a procedure called gauge fixing) a universal gate set in a fault-tolerant manner. This last property would preclude the need for procedures like magic state distillation, which are currently foreseen to take up the vast majority of computing resources in other fault-tolerant quantum computation schemes.

We remark that entangling operations of surface code logical qubits encoded in two different qubit modules can be performed by shuttling only the qubits at the edges to the other qubit modules [67], see Fig. 7, and subsequently returning the qubits to the original module. This avoids the necessity to shuttle all qubits in one module to the next in order to perform two qubit gates between logical qubits.

We could foresee lower performance regimes or faulty qubits on the chip, for example due to the qubit-to-qubit variation induced by the ESR stripline pair. One way to address faulty sites within one qubit module would be to change the actual quantum error correcting code to encode one (or more) logical qubits with fewer physical qubits using the remaining qubits in the vicinity [67]. Yet, it is clear that this introduces inhomogeneity in the classical control requirements of the individual modules, and greatly complicates two-qubits gates between two logical qubits, as they are now encoded using different codes. Depending on the fidelity of the long-distance shuttling operations in the fabricated devices, however, another path could be to turn off qubit modules completely if the noise exceeds a certain threshold. As a consequence, we may need to shuttle qubits over longer distances in order to perform two-qubit operations on logical qubits, but would have the ability to select the desired good qubit modules. This is particularly promising in this architecture, given the ability to shuttle fast and with high fidelity.

A particular challenge is to map quantum circuits to our architecture. For this, a variety of classical methods exist. To gain maximum advantage of the ability to shuttle qubits, the long-distance shuttling operations are ideally fast compared to general gate speeds. In this case, the architecture becomes virtually non-planar which can yield significant savings in overhead [68].

While many traditional quantum algorithms such as Shor's factoring algorithm require a large number of qubits, few qubit applications are slowly beginning to emerge. In recent years, interest in electronic structure quantum simulation has culminated in small-scale experimental implementations [69,70]. Larger simulation algorithms will have to deal with entangling large amounts of qubits along certain paths across the device, as introduced by the standard mapping of second quantization. The switching to different mappings on the other hand [71–73] reduces the amount of gates but does not solve the connectivity problems, which the proposed architecture is a promising candidate to tackle. Shuttling and the native $\sqrt{\text{SWAP}}$ gates might also be used to move certain auxiliary qubits around, which allows for significant decreases in depth of the resulting quantum circuit [74].

Upscaling towards the numbers of qubits required for these algorithms, including few qubit applications, represents a formidable challenge. However, we envision that the proposed architecture based on shared control and flexible qubit shuttling can provide a unique shortcut towards large-scale quantum computation.

# ACKNOWLEDGMENTS

M.V. acknowledges support by the Netherlands Organization of Scientific Research (NWO) VIDI program. JH and SW are funded by an NWO Vidi grant, an ERC Starting Grant and STW Netherlands. MS is funded by NWO/OCW and an ERC synergy grant. L.M.K.V. acknowledges support by the Netherlands Organization of Scientific Research (NWO) VICI program.

# Supplementary Material:
# A Crossbar Network for Silicon Quantum Dot Qubits

R. Li[1,2], L. Petit[1,2], D.P. Franke[1,2], J.P. Dehollain[1,2], J. Helsen[1], M. Steudtner[3,1], N.K. Thomas[4], Z.R. Yoscovits[4], K.J. Singh[4], S. Wehner[1], L.M.K. Vandersypen[1,2,4], J.S. Clarke[4], and M. Veldhorst[1,2]*

[1]*QuTech, Delft University of Technology, P.O. Box 5046, 2600 GA Delft, The Netherlands.*
[2]*Kavli Institute of Nanoscience, Delft University of Technology, P.O. Box 5046, 2600 GA Delft, The Netherlands.*
[3]*Instituut-Lorentz, Universiteit Leiden, P.O. Box 9506, 2300 RA Leiden, The Netherlands.*
[4]*Components Research, Intel Corporation, 2501 NW 229th Ave, Hillsboro, OR 97124, USA.*
*email address: m.veldhorst@tudelft.nl

## Section 1. Tolerance to quantum dot inhomogeneity

In this section, we discuss the required homogeneity for the shared gate control. Firstly, we estimate the upper bound of the inter-dot tunnel coupling when the tunnel barriers are set to the off-state. Finite tunnel coupling in the off-state can result in unwanted shuttling of electrons. These shuttle processes need to be error corrected in quantum algorithms. Here, we consider the surface code operation [5], to estimate the error correction cycle. Importantly, undesired shuttle events could occur at any coordinate of the quantum module. Consequently, we need to take into account the idle qubits in each step and consider the complete cycle time of the qubit module. We target for a shuttle-error rate below 0.1% in a complete error-correction cycle. Most errors are expected during the readout of the measurement qubits, due to speed and pulsing requirements. Row-controlled PSB can be performed at a frequency of 3 MHz, such that spin-to-charge conversion on a 50 × 50 qubit module size can be within 10 μs. In order to achieve a 0.1% error rate, electrons should have a shuttle rate at least $10^3$ smaller than the cycle time, and we require $t_{0,\text{off}} < 10$ Hz in the off-state.

Now we discuss the tunneling rate range when the barriers are set to the on-state. Since the desired qubit shuttling rate is at least 1 GHz, we require a minimum coupling of $t_{0,\text{on}} > 10$ GHz. The upper bound in the tunnel rate follows from the requirement that a larger tunneling rate needs a larger detuning to prevent charge state mixing between different quantum dots (see Section 5 for details). While in principle larger voltage pulses may enable this, larger detuning could lead to overhead in operations and bring down undesired higher orbital levels. Therefore, we require the tunneling rate variation to be within one order of magnitude with an upper bound for the tunneling rate $t_{0,\text{on}} < 100$ GHz.

Next to the tunnel coupling, the uniformity requirement on the quantum dot chemical potential $\Delta\mu$ are also crucial. When $\Delta\mu < E_C$, where $E_C$ denotes the charging energy, all the quantum dots under the same $QL$ can be controlled to have the same charge state. Importantly, this qubit occupation configuration is the ground state even when the inter-dot tunnel couplings are set to on. We envision this to be beneficial in order to correct shuttle errors. Although the qubit configuration in Fig. 1(b) can also be achieved with $\Delta\mu > E_C$, by sequentially shuttling the qubits to the desired sites, we point out that lower uniformity demands higher detuning for compensation. This could slow down the gate pulsing speed and hence the overall operation rate, and impractical voltage pulses may be required.

The regime $\Delta\mu < E_C$ is also important for operations like parallel PSB, as shown in the main text, Fig. 5(c). When variations in $\mu$ are large, electrons such as the one labelled 2 in the figure may shuttle to an adjacent column. While such errors could be corrected by another phase-controlled shuttle, we envision this to be impractical and providing a significant overhead. First, it will significantly slow down pulses to ensure adiabaticity. Second, it will require large voltage pulses in order to overcome the variations. Third, after a PSB step, a shuttle step has to be implemented purely for pulsing back electrons to their targeted positions. Instead, when $\Delta\mu < E_C$, these requirements are avoided altogether, since the spread in chemical potential energies is smaller than the charging energy that separates different electron occupations, such that electrons will remain in their targeted positions.

In conclusion, we require the chemical potential variations $\Delta\mu < E_C$, and the tunnel coupling $t_{\text{off}} < 10$ Hz and 10 GHz $< t_{\text{on}} < 100$ GHz.

# Section 2. Column-by-column alternating static magnetic field

In this section, we estimate the column-by-column alternating magnetic field generated by the constant current through the *CL*. The *CL* have width $w_{CL}$ = 30 nm and height $h_{CL}$ = 60 nm. Since the effective superconducting penetration depth $\lambda_{eff}$ can be much larger then these dimensions, we assume a uniform current density through the *CL*. The resulting magnetic field is calculated along the row direction and in the plane 20 nm below the bottom of the *CL* grids, where the quantum dot qubits are located. For generality, we take the *CL*s infinite in length and calculate the field strength via the Biot-Savart Law. We note that this assumption will hold for the large arrays assumed here, although the edges will require corrections. The rectangle shape of *CL* is approximated by dividing it into square node points that are uniformly spaced in the rectangle with equal currents (30 x 60 nodes). In total, we calculate 40 *CL*s and convert the field to resonance frequency using $g$=2, and plot the middle region as shown in Fig. 2(d) of the main text.

We further estimate the local resonance frequency variation, $\Delta v_0$, due to imperfect device fabrication or inhomogeneity in quantum dot position and size. In Fig. S5(a-d), we show the influence of a deviation of a certain geometry in *CL* gate on $\Delta v_0$, by comparing it to ideal *CL* gates. The left panels shows the color-coded relative resonance frequency error with respect to the designed value along the row direction (x-axis) for different fabrication errors (y-axis). Although we find that $\Delta v_0$ can be significantly, the maximum is in between the quantum dots and the amplitude is strongly reduced at the center of the qubit location. To estimate the influence on the quantum dot position, we calculate the average resonance frequency error, $\Delta v_{0,\text{ave}}$, for different dot sizes, as shown in the right panels. By comparing the various results, we can see that with the same absolute fabrication error, the offset of the *CL* gate height has the strongest effect, especially for the misalignment on the bottom side [Fig. S5(c)]. We envision therefore that it will be crucial to choose an integration scheme that minimizes the roughness under the gate area. The influence of quantum dot geometry on the field is relatively weak, as shown in Fig. S5(e)-S1(f), since the out-of-plane field is not sensitive to the dot position as shown in Fig. 2(d) of the main text.

In Fig. S6, we plot the sum of all the errors presented in Fig. S5 with a dot size of 20 nm. For a 1 nm error in fabrication or dot geometry homogeneity, the maximum change in magnetic field $\delta v_{fab}$ = 100 kHz (Note that this number is not sensitive to the quantum dot size. The limiting factor in $\Delta v_0$ is *CL* height, which shows weak dependence on the quantum dot size).

We estimate the magnetic field at the dot sites by taking linear averages for different dot sizes. The real electron wave function will have a different distribution, and will distribute more in the middle than the edge. Since we find that the center is most insensitive to fabrication errors, the results shown here can be taken as the upper bound on requirements. This also becomes clear from Fig. S5, where we see that larger dot sizes will generally contribute to higher magnetic field deviations. In addition, if the self-correlation length of the geometry error is smaller than the effective span of the electron wave function, the overall resonance frequency variation will be even smaller. Therefore, we estimate that for 1 nm root-mean-square (rms) variation in the gate geometry, the qubit to qubit resonance frequency variation is in the range of $\delta v_{fab}$ = 100 kHz. While these numbers are certainly challenging, industrial fabrication has pushed uniformity to the limit, such that alignments with nm resolution are possible [30,31].

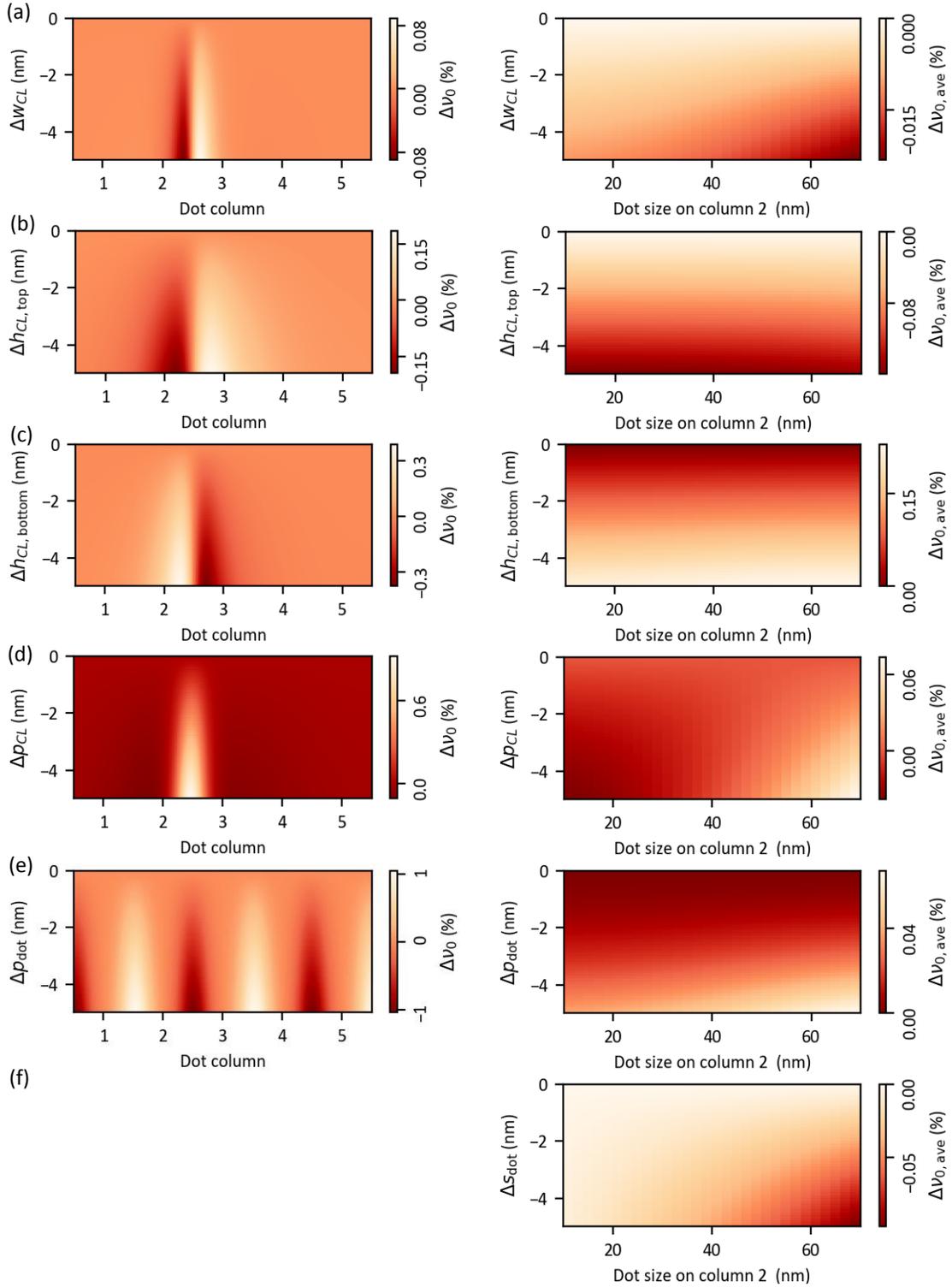

Fig. S5: Impact of misalignment and errors in gate and dot dimensions. Error in $\Delta v_0$ for errors in (a) the CL gate width $\Delta w_{CL}$, (b) the CL gate top location $\Delta h_{CL,top}$, (c) the CL gate bottom location $\Delta h_{CL,top}$, (d) the *CL* gate lateral location $\Delta p_{CL}$, (e) the dot location $\Delta p_{dot}$, and (f) the dot size $\Delta s_{dot}$. The left panels show the errors normalized with respect to the targeted perpendicular magnetic field. The right panels show the averaged errors, $\Delta v_{0,ave}$, taking into account the finite size of the quantum dots.

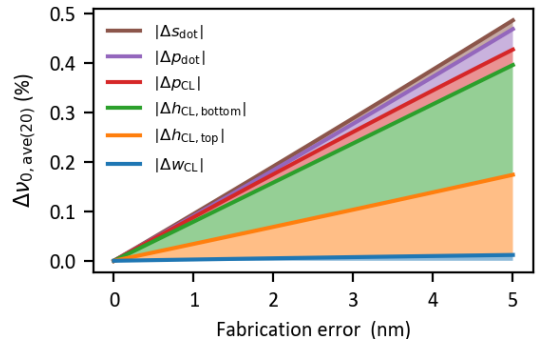

Fig. S6: Overall resonance frequency error as a function of fabrication error. The $\nu_0$ error is calculated based on a dot size of 20 nm. Different resonance frequency errors are stacked on top of each other to show the worst case scenario assuming the same deviation in gate or dot geometry. Controlling the vertical dimension is most critical.

# Section 3. Inhomogeneity of the ESR stripline

Simultaneous qubit control requires the amplitude of the spin-resonant magnetic field to be highly homogenous, such that all resonant qubits respond with the same Rabi frequency. In order to estimate the homogeneity and optimize the design of our stripline, we turn to the Microwave Studio simulation package from Computer Simulation Technology (CST-MWS) [41]. With this 3D simulator of high-frequency devices we can create a 3D model of our stripline structure, define ports for excitations and solve Maxwell's equations over a finite-element mesh of our model.

Fig. S7(a) shows a schematic of the stripline model we have designed and simulated. A qubit module includes a pair of narrow striplines placed above the qubit plane. Current flowing through the striplines generates a magnetic field that wraps around the striplines. Therefore, the qubit module experiences the in-plane component of this field. A stripline pair is chosen because we can obtain the same homogeneity as for the case of a single but wider stripline, while significantly less current is required. The stripline pair is furthermore simple in design. The model consists of a lossless silicon substrate with superconducting striplines on the surface. CST-MWS models these lines with a frequency dependent surface impedance and equal penetration depth $\lambda$ over all frequencies. The striplines fan out to a short-circuited coplanar waveguide structure, similar to those described in reference [42]. We used the frequency domain solver and analyzed our results at $v_0 = 1$ GHz.

Using the parametric optimization function built into the CST-MWS simulator, we run a sweep of simulations to optimize for field homogeneity. Here, we vary the stripline width ($w_{stripline}$), the pitch between the striplines ($d_{stripline}$), and the separation between the striplines and the qubit module ($h_{stripline}$). Fig. S7(b) shows the plots of the homogeneity along one axis of the qubit module for parameter combinations we tested. We find RF field inhomogeneity across the 2D array $\delta v_{Stripline} < 2$ %, for $w_{stripline} = 2$ μm, $d_{stripline} = 6$ μm, $h_{stripline} = 4$ μm.

We have analyzed a range of superconductor penetration depths (with $\lambda$ ranging from 0.5 μm to 5 μm) and found only minor variations, demonstrating the robustness of our design and enabling to use a range of superconducting materials and film-thicknesses for the stripline.

Additionally, we can extract the current density along our striplines by integrating the magnetic field along a cross-section of the stripline, which is readily available as part of the simulation results.

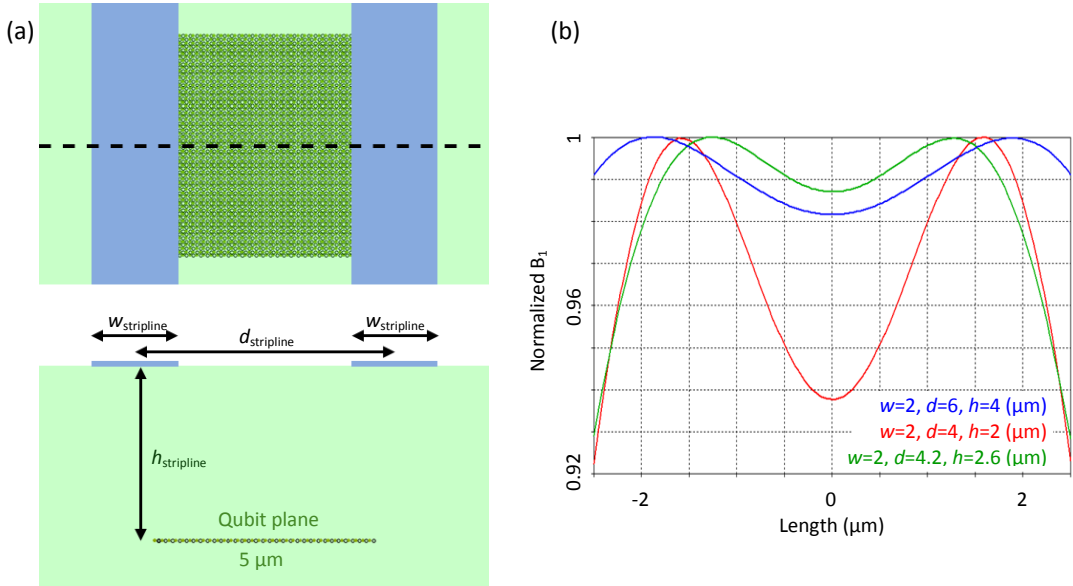

Fig. S7: Stripline schematic and simulation results. (a) Top image shows a top view of the stripline design with the superconducting metal strips in blue, the silicon substrate in green and the qubits represented as small circles. Bottom image is a cross-section along the dashed line, showing the qubits directly under the stripline pair. The relevant dimensions for the design are labelled. (b) Field homogeneity for different stripline design dimensions. We configured the optimization algorithm from CST to test different design dimensions to maximize the homogeneity. The figure shows some of the results obtained, with the optimal result found in blue.

# Section 4. Grape pulse for spin rotation

A crucial point in the single qubit manipulation across the 2D array is that the applied ESR pulse can address the qubits with the larger (smaller) $v_0$ without effecting the qubit with the smaller (larger) $v_0$. At the same time, however, it needs to tolerate variations in the static field and in the ESR field. As discussed in the main text and Supplementary Materials Section 3, the ESR field inhomogeneities can be engineered to be $\delta v_{Stripline} < 2\%$. The variation in the qubit resonance frequency are estimated to be $\delta v_0 \sim 150$ KHz for a frequency difference of 10 MHz between the columns. In Fig. S8(a) and S4(b) we show the gate fidelity of the targeted qubits ($v_0 = 105$ MHz) and idle qubits ($v_0 = 95$ MHz) as a function of variations in $v_0$ and $v_1$ when naively applying a 1 MHz square ESR pulse. The target gate for the 105 MHz qubit is a $\pi/2$ rotation, while for the 95 MHz qubit it is the identity operator. The resonant qubit is rotated with >99.9% fidelity with tolerances for detuning around 100 kHz in $v_0$ and 3% in $v_1$. However, the off-resonant qubit does not achieve the targeted fidelity even for null detuning.

For these reasons, we have made use of numerical techniques to identify a composite pulse that can meet our requirements. The scheme we adopted is the Gradient Ascent Pulse Engineering (GRAPE) [47]. In the algorithm, the time evolution of the system is split in small timeslots in which the amplitude of the pulse is assumed to be constant. For each timeslot, the amplitude is then optimized using standard multi-variable optimization methods in order to maximize the overlap between the actual gate and the target gate. Since the goal is to obtain a selective pulse tolerant to detuning in resonance frequency, we evaluate the simulation result as an average of the fidelities of four qubits: two qubits with frequencies $105 \pm 0.1$ MHz targeted to be on resonance and two qubits with resonant frequencies $95 \pm 0.1$ MHz targeted to be off resonance. In the simulation, it is important to limit the number of qubits, since increasing the search space can also increase the number of local maxima with insufficient high fidelity such that the algorithm is incapable of solving the problem. We note that because of the low Rabi frequency compared to the Larmor frequency, we do not take the rotating wave approximation.

In Fig. S8(c) [(d)] we show the average gate fidelity for a $\pi/2$ [null] rotation using the optimized GRAPE pulse for the respective qubits. Fidelities beyond 99.9% can be achieved up to 300 kHz in $v_0$ and over 3 % in $v_1$. The gate can be executed in the same time as a 2 MHz Rabi pulse, i.e. in 250 ns. This comes at the cost of a slightly larger RMS amplitude (1.1 MHz compared to 0.7 MHz) with a maximum peak of ~3 MHz.

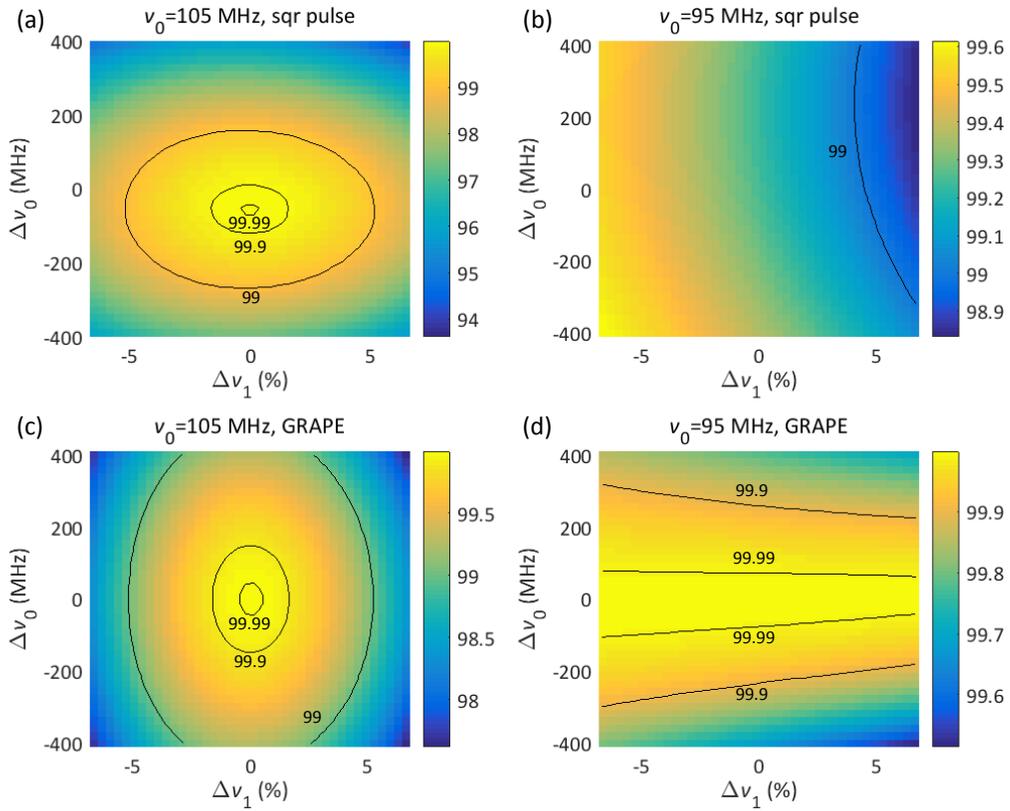

Fig. S8: GRAPE pulse optimization for high fidelity single qubit gates. Gate fidelity for a square pulse (a) and (b) and for an optimized GRAPE pulse (c) and (d), targeting in both cases a π/2-rotation and identity-gate on the qubits with higher and lower resonant frequency, respectively. The pulse is shown as a function of variations in resonance frequency $v_0$ and ESR field $v_1$ to test the robustness of the pulse. For the optimized GRAPE pulse we can find that the fidelity can be readily above 99.9 %.

# Section 5. Shuttling fidelity

In this section, we estimate the shuttling fidelity considering various inter-dot energy detunings, tunnel couplings and shuttling speeds. Firstly, we calculate the required detuning to isolate the qubits. When the tunnel barrier is set to on by *CL* or *RL*, the tunnel coupling $t_0$ mixes different charge states also around idle qubits. Here, we consider the lower bound, which is the situation where the tunneling barrier between an idle qubit and an empty dot is turned on diabatically. Then the charge state, say, (1,0), will process around the eigenstate. To maintain the minimum (1,0) fraction higher than *F*, the inter-dot energy detuning between the idle qubit and the empty dot needs to be larger than $\varepsilon_F = t_0|2Fh - 1|/\sqrt{Fh - Fh^2}$, with the (1,0) fraction in the eigenstate $Fh = \sqrt{(F+1)/2}$. We find consequently for the case $t_0 = 100$ GHz and charge fraction $F = 99.9\%$ that the minimum required detuning is $\varepsilon_F = 26$ meV.

To separate different charge state during the shuttling process, we also consider the effect of $\Delta\mu$. As shown in Fig. S9(a), the clearance for each state is $\Delta\mu + \varepsilon_F$. When shuttling a single qubit as shown in Fig. 3(b) and (c) in the main text, their neighboring qubits controlled by *QL3* should not be affected. Fig. S5(a) shows a scheme that with high *QL* pulsing amplitude of $2(\Delta\mu + \varepsilon_F)$, the *QL3* qubit remains with a negative detuning. Fig. S5(b) shows an alternative scheme where the pulsing amplitude can be halved by applying a pulse on *QL3* to compensate the change in detuning. The choice between fewer control signals or faster operation together with smaller pulsing amplitude can be made based on the physical qubit properties and control circuitry specifications.

We now discuss a linear and adiabatic pulsing scheme. We describe the non-adiabaticity by the Landau–Zener formula, $P_n = \exp(-4\pi^2 t_0^2 \Delta t/\Delta\varepsilon)$, where $\Delta t$ is the shuttling time and $\Delta\varepsilon$ is the total detuning sweep range. To implement parallel linear shuttling scheme, we need to account for the largest pulsing amplitude with the smallest tunneling rate in order to tolerate quantum dot variations. The tunneling rate of $t_{0,max} = 100$ GHz requries $\varepsilon_{F,\max} = 26$ meV (although higher $t_0$ allows faster shuttling, the required $\varepsilon_F$ could be impractical to achieve). Combining this with the smallest tunneling rate $t_{0,min} = 10$ GHz and $P_n = 10^{-3}$, we find the line-by-line parallel shuttling rate $f_{parallel} \sim 45$ MHz.

In order to pulse faster, we have developed a new protocol. The results are shown in Fig. S5(c) and (d). In this protocol we first reduce the detuning while the tunnel coupling is off, then we turn on the coupling adiabatically (0.2 ns), apply the linear adiabatic detuning pulse (0.6ns), and finally we turn of the coupling adiabatically (0.2 ns) and set the detuning back to the idle value; corresponding to 1 GHz shuttling rate. When the tunnel barrier is set to off, there is no hard boundary on the detuning pulsing speed. The tunnel barrier can be adiabatically turned on with a timescale below nano-seconds, since the detuning energy is generally much larger than the tunnel coupling. As marked by the black contour line in Fig. S5(c), higher than 99.9% shuttling fidelity can be achieved requiring a minimal tunneling rate $t_{0,min} > 10$ GHz, and it does not pose an upper bound to $t_{0,max}$. We note that while this example demonstrates proof-of-principle further optimization is possible [51].

For global shuttling [Fig. 6(a)] we need to take into account the chemical potentials variations between different dots, and $\Delta\varepsilon > 2\Delta\mu$, such that the pulsing amplitude is larger than the variation. We can use this requirement together with the result shown in Fig. S5(c) to find the associated tunnel coupling for a 1 GHz shuttling rate. Considering $\Delta\mu = 2$ meV, we find $t_{0,min} > 20$ GHz.

Now we consider the global RF-dispersive based charge readout using the frequency multiplexing scheme as shown in Fig. 6(b). Here, we focus on the quantum capacitance $C_q$, which affect the signal strength, and the full width at half maximum (FWHM), which affect the simultaneous measurement range. In the low temperature high $t_0$ limit, FWHM $\sim 3t_0$; and at zero detuning, $C_{q0} \sim (e\alpha')^2/4t_0$, where $\alpha'$ is the lever-arm different between two dots [59]. For $\Delta\mu \sim 2$ meV and $t_{0,min} = 10$ GHz, we need $\sim 16$ measurement cycles to cover $\Delta\mu$ with FWHM. For $t_{0,min} = 10$ GHz, the minimal quantum capacitance $C_{q0}/2 \sim 19$ aF. For $t_{0,max} = 100$ GHz, $C_{q0} \sim 3.8$ aF, but the FWHM is much wider and it can be integrated over several measurement cycles.

In the last part of this section, we discuss the charge pulsing for PSB spin-to-charge conversion [results shown in Fig. S5(e) and (f)] and assume long spin lifetimes. The first step involves turning on $t_0$. This step is limited in speed due to the small direct coupling between the $|\uparrow\downarrow\rangle$ and $|\downarrow\uparrow\rangle$ components. Next, we apply a linear detuning pulse to shift the lower energy eigenstate, e.g. the $|\uparrow\downarrow\rangle$-like state, into S(0,2). Now, because of the small direct coupling, the pulsing needs to be only adiabatic with $t_0$ and not to the Zeeman energy difference. Finally, we turn off $t_0$. This step can also be fast as there is no other states close to S(0,2) at positive detuning. A high-fidelity 3 MHz PSB spin-to-charge conversion rate is achieved by turning on $t_0$ in 200 ns, followed by linear sweeping the detuning in 110 ns, and turning off $t_0$ in 20 ns. Fig. S5(e) shows the operation fidelity as a function of $t_0$ and $\varepsilon$, where the black contour line denote the region where the fidelity is beyond 99.9%.

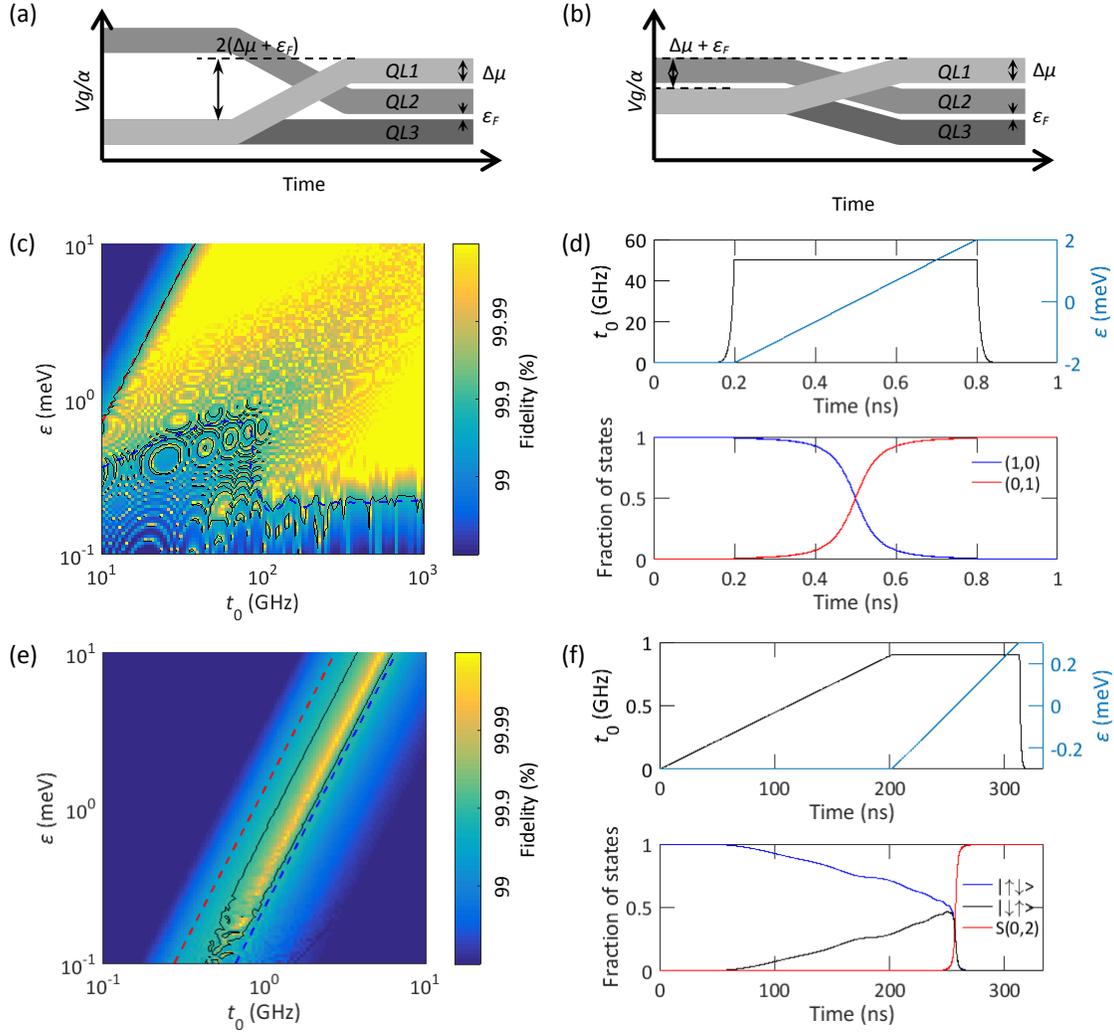

Fig. S9: Charge shuttling process. (a) and (b) Qubit shuttling scheme for the same operation as Fig. 3(b) or 3(c) in the main text. (a) *QL1* and *QL2* on top of the shuttling sites are pulsed with a larger amplitude such that the qubit under *QL3* is not affected. (b) *QL3* is also pulsed to compensate the reduced detuning, and the pulsing amplitude is reduced. (c) The fidelity of a three-step shuttling with an operation speed of 1 GHz. The first step is turning on the inter-dot tunnel coupling from 1 Hz to $t_0$ in 0.2 ns. The second step is a linear sweep of the detuning from $-\varepsilon$ to $\varepsilon$ in 0.6 ns. The last step involves turning off the tunnel coupling in 0.2 ns. Here, we consider a fast, linear control voltage on the barrier gate and approximate the tunneling rate change by an exponential scale. (d) An example shuttling process with $t_0 = 50$ GHz and $\varepsilon = 2$ meV. (e) The fidelity of a three-step PSB spin-to-charge conversion with an operation speed of 3 MHz. The first step is turning on the inter-dot tunnel coupling linearly from 1 Hz to to $t_0$ in ~ 200 ns. The second step is linearly sweep the detuning from $-\varepsilon$ to $\varepsilon$ in ~ 110 ns. The last step is turning off the tunnel coupling back to 1 Hz in ~20 ns. (f) An example PSB process with $t_0 = 1$ GHz and $\varepsilon = 0.3$ meV, where $|\uparrow\downarrow\rangle$ has a lower energy. In (c) and (e), black contour lines denote the 99.9% fidelity-threshold, red dashed lines correspond to a non-adiabatic probability of $10^{-3}$ from the Landau–Zener formula during the detuning sweep, and the blue dashed contour lines denote the 99.9% fidelity-threshold during tunnel coupling control.

## Section 6. Pauli spin blockade spin to charge conversion with ancillary qubit in the spin up state

In this section, we explain how to implement PSB readout with the ancillary qubit in the spin up state. This complements the protocol with the ancillary qubit in the spin down state, as described in the main text and visualized in Fig. 4(b). In this protocol, the spin up ancillary qubit is located in the column with the smaller magnetic field (red column) at the starting of the PSB process. Consequently, the qubit for readout is pulsed to the ancillary qubit site, as shown in Fig. S10(a) denoted by step (i). We note that an alternative sequence is possible as well, obtained by reversing the pulsing direction. As shown in Fig. S10(b), if the target qubit is in the spin up state, it will remain in the (1,1) charge state. In contrast, if the state is spin down, it will move to the S(0,2) state. Consequently, the charge occupation can be readout with the gate-based dispersive readout as described in the main text.

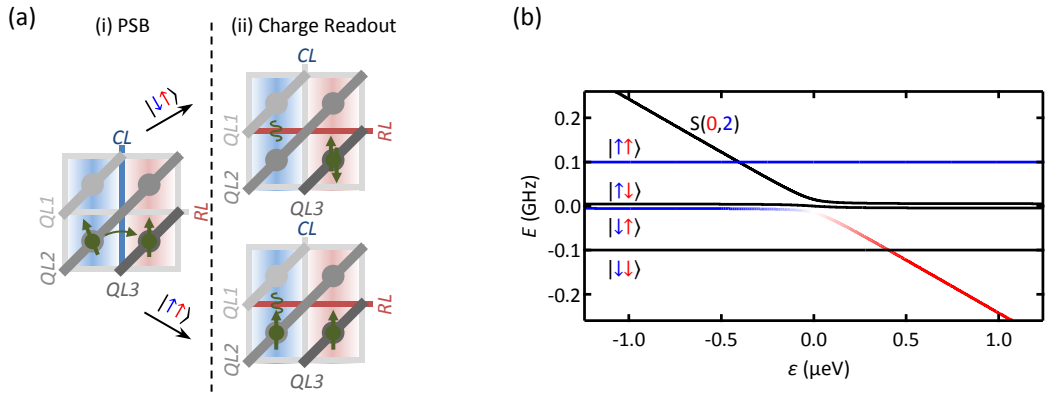

Fig. S10: Scheme for Pauli spin blockade spin to charge conversion with ancillary qubit in the spin up state. (a) Schematic for (i) the PSB and for (ii) the charge readout process. (b) Double dot energy diagram. The ancillary qubit with smaller Zeeman energy (on the red column) is tuned to the up state. The readout state are the spin states of the qubit with the larger Zeeman energy. By increasing adiabatically the detuning beyond the anticrossing location, the individual spin states are projected to a singlet (denoted by the blue to red line) or triplet state (blue line). The resulting difference in charge state can be consequently measured as described in the main text.

# Section 7. Shuttling bus for 2D array module

In this section, we discuss briefly a direction how to shuttle qubits between different qubit modules as a means towards truly large-scale quantum computation. The architecture is shown in Fig. S11. With this approach it is possible to shuttle individual or complete arrays of qubits using the shuttling gates. In this implementation, the shuttling gates have the same geometry as the *CL*s and *RL*s. Fig. S11(i) denotes the starting position of the qubits for a possible shuttle sequence. In Fig. S11(ii), we lower the tunnel barriers in the forward direction of the qubits with label 1. Next, we raise the tunnel barriers in the backward direction, and hence they are shuttled forward. In Fig. S11(iii), we repeat the same process as step (ii), but now for the qubits with label 2. After these steps, the qubit array is shuttled forward by one dot site as shown in Fig. S11(iv). In this scheme, the shuttling gates are grouped in four, where the gates in each group perform similar operations. Consequently, all the gates belonging to the same group can be connected together to further reduce the number of wires interfacing to external electronics.

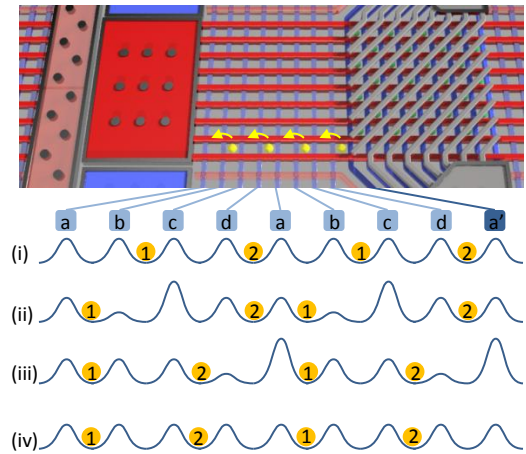

Fig. S11: Connecting qubit modules. The shuttling highways need fewer gate lines as compared to the qubit module. This limits functionality but does allow to shuttle qubits between different qubit modules. In addition, the gate lines can be grouped, such that space becomes available for interconnects or local electronics. The bottom section of this figure schematically shows a particular shuttle scheme, where a column of qubits is shuttled by one dot site by advancing from steps (i) to (iv).